\documentclass[print,twocolumn]{revtex4}
\usepackage{amssymb}
\usepackage{amsmath}
\usepackage{graphicx}
\usepackage{dcolumn}
\usepackage{bm}
\usepackage{txfonts}
\usepackage{appendix}
\usepackage[dvips]{color}
\usepackage[colorlinks, linkcolor=blue, anchorcolor=blue, citecolor=blue, urlcolor=blue ]{hyperref}

\begin{document}
\title{High rectifying performance of heterojunctions with interface between armchair C$_3$N nanoribbons with and without edge H-passivation}

\author{Jie Zhang$^1$}
\author{Wence Ding$^1$}
\author{Xiaobo Li$^2$}
\email{xiaoboli2010@hutb.edu.cn}
\author{Guanghui Zhou$^1$}
\email{ghzhou@hunnu.edu.cn}

\affiliation{$^1$Department of Physics, Key Laboratory for Low-Dimensional Structures and Quantum Manipulation (Ministry of Education), Hunan Normal University, Changsha 410081, China}

\affiliation{$^2$Department of Applied Physics, School of Science, and Key Laboratory of Hunan Province for Statistical Learning and Intelligent Computation, Hunan University of Technology and Business, Changsha 410205, China}

\begin{abstract}
Two-dimensional polyaniline with C$_3$N stoichiometry, is a newly fabricated layered material that has been expected to possess fascinating electronic, thermal, mechanical and chemical properties. The nature of its counterpart nano-ribbons/structures offering even more tunability in property because of the unique quantum confinement and edge effect, however, has not been revealed sufficiently. Here, using the first-principles calculation based on density functional theory and nonequilibrium Green's function technique, we first perform a study on the electron band structure of armchair C$_3$N nanoribbons (AC$_3$NNRs) without and with H-passivation. The calculated results show that the pristine AC$_3$NNRs are metal, while the H-passivated ones are either direct or indirect band gap semiconductors depending on the detailed edge atomic configurations. Then we propose a lateral planar homogenous junction with an interface between the pristine and H-passivated AC$_3$NNRs, in which forms a Schottky-like barrier. Interestingly, our further transport calculation demonstrates that this AC$_3$NNRs-based heterojunction exhibits a good rectification behavior. In specification, the average rectification ratio (RR) can reach up to $10^3$ in the bias regime from 0.2 to 0.4 V. Particularly,  extending the length of semiconductor part in the heterojunction leads to the decrease of the current through the junction, but the RR can be enlarged obviously. The average RR increases to the order of $10^4$ in the bias from 0.25 to 0.40 V, with the boosted maximum up to $10^5$ at 0.35 V. The findings of this work may be serviceable for the design of functional nanodevices based on AC$_3$NNRs in the future.
\end{abstract}

\maketitle

\section{Introduction}
Two-dimensional (2D) nanomaterials, such as graphene \cite{Novoselov1,Novoselov2,Castro Neto}, transition-metal dichalcogenides (TMDs) \cite{Strano,Chhowalla} and phosphorene \cite{Li,Liu,Qiao,Xia}, play an important role in the development of nanotechnology. Particularly, various interesting electronic transport properties have been reported for nanodevices based on 2D nanomaterials, such as rectification \cite{Baugher,Cao1}, negative differential resistance (NDR) \cite{Habib,Wang,H. Ren}, switching \cite{Wan1,Moon}, spin filtering \cite{Karpan,Wan2,Cao2} and field-effect transistors \cite{Jiang,Cao3,Z. X. Yang}. However, the gapless of graphene \cite{Castro Neto}, the instability of phosphorene \cite{Joshua,Li} and the low carrier mobility of MoS$_2$ \cite{Desai,Yu} make some limitations to their usage in electronic devices. People expect to, therefore, find a novel 2D material without these shortages. Fortunately, the 2D layered material polyaniline C$_3$N has recently been successfully produced by using polymerization of 2,3-diaminophenazine method \cite{Yang}. In comparison with graphene, C$_3$N has higher stiffness, stability and on-off current ratio ($\sim$10$^{10}$). Moreover, the monolayer C$_3$N is a semiconductor with an indirect band gap 0.39 eV \cite{Yang,A. Bafekry}, which may be unsuitable for electronic device applications \cite{B. Mortazavi,Y. Ren}. Therefore, under this motivation, we pay particular attention to the the possibility of indirect-direct gap transition in C$_3$N nanostructures \cite{A. Bafekry,B. Mortazavi,Y. Ren}.

As is well known that the electronic property of a 2D material nanoribbon depend on its width and edge geometry. For example, an intrinsic zigzag-edge graphene nanoribbon (ZGNR) has an edge state and exhibits metallicity; while the bandgap of the armchair-edge graphene nanoribbon (AGNR) decreases with the increase of width and exhibits the rule of 3$n$, 3$n$+1, 3$n$+2 ($n$ is the number of the carbon chains along the width of the ribbon) \cite{K. Nakada,Y.-W. Son,V. Barone,Zhou}. Further, whether a phosphorene nanoribbon is semiconductor or metal heavily depending on edge modification for both edge types \cite{Vy Tran,A. Carvalho,W. F. Li}. For a zigzag-edge MoS$_2$ nanoribbon (MoS$_2$NR), moreover, it may exhibit half-metallic, metallic or semiconducting characteristic depending on different edge reconstruction or hydrogen (H)-passivation \cite{Y. F. Li}. More interestingly, the armchair-edge MoS$_2$NR shows nonmagnetic semiconducting behavior, and the bandgap has a tendency to oscillate with width \cite{Y. F. Li}. These fascinating physical properties promote the development of field-effect transistors \cite{Q. Y. Wu,S. Fathipour,V. Passi}.

The C$_3$N nanoribbons, however, may offer more tunability in electronic property because of the unique edge effect and quantum confinement. Several studies have shown that the band structure and gap size of C$_3$N nanoribbons can also be modulated by the edge termination and ribbon width \cite{A. Bafekry,Y. Ren}. Moreover, the H-passivated C$_3$N nanoribbons have attracted particular attention because of their stability. For example, Xia \emph{et al}. \cite{Congxin Xia} reported that the H-passivated zigzag C$_3$N nanoribbons (ZC$_3$NNRs) with edges composed of both C and N atoms are semiconductors, and their bandgaps decrease with the increase of ribbon width. When the width is large enough, the band gap tends to a constant, which is the value of the C$_3$N nanosheet. Further, Bafekry \emph{et al}. \cite{A. Bafekry} also demonstrated that the band gaps of the H-passivated armchair C$_3$N nanoribbons (AC$_3$NNRs) converge to different values as the width increasing, neither of which are that of the C$_3$N nanosheet. This is because the conduction band minimum (CBM) and valence band maximum (VBM) are determined by the edge geometry \cite{A. Bafekry}. However, the electronic and transport properties of C$_3$NNRs-based lateral heterojunctions have rarely been addressed.

To know whether the C$_3$NNRs have potential in practical application, in this work we investigate the H-passivation effects on the electric and transport properties of AC$_3$NNR and its related Schottky-like junction. It is worth noting that, unlike the pure-carbon graphene, the AC$_3$NNRs have two edge morphologies, i.e., all-carbon or C-N mixed morphology \cite{A. Bafekry,Y. Ren}. Therefore, for a AC$_3$NNR-based heterojunction, in addition to the H-passivation, the edge morphology is also an essential factor to be considered. Using the density functional theory (DFT), our calculation concludes that the variation either in edge morphology or the H-passivation can effectively regulate the electronic structure of the AC$_3$NNRs, thereby affecting  the nature of the related heterojunctions. Further, we construct three Schottky heterojunction devices composed of the pristine AC$_3$NNRs with different edges and the corresponding H-passivated AC$_3$NNRs. Since the pristine AC$_3$NNRs are metal and the H-passivated AC$_3$NNRs are semiconductor, a metal-semiconductor planar junction forms a Schottky interface barrier. Interestingly, our further transport calculation demonstrates that this AC$_3$NNRs-based heterojunction exhibits a good rectification behavior. In specification, the average rectification ratio (RR) can reach up to $10^3$ in the bias regime from 0.2 to 0.4 V. Particularly,  extending the length of semiconductor part in the heterojunction leads to the decrease of the current through the junction, but the RR can be enlarged obviously. The average RR increases to the order of $10^4$ in the bias from 0.25 to 0.40 V, with the boosted maximum up to $10^5$ at 0.35 V. The findings of this work may be serviceable for the design of functional nanodevices based on AC$_3$NNRs in the future.

This paper is organized as follows. Sec. II provides the description of the AC$_3$NNRs and their based heterojunctions as well as the computational details. In Sec. III, we present the calculated results and the discussions with physical mechanism explanations. Finally, we give a brief summary in Sec. IV.

\section{Models and Computational methods}
To explore the electronic and transport properties of AC$_3$NNRs, firstly, we justify the structural parameters of a C$_3$N nanosheet. The optimized C$_3$N unit cell has a planar hexagonal lattice symmetry and the lattice parameter is 4.87 {\AA}, including two N atoms and six C atoms. The previous calculated band structure has shown that the C$_3$N nanosheet is an indirect band gap semiconductor with the gap value of 0.39 eV, and the lattice parameter is 4.862 {\AA} \cite{Yang,A. Bafekry,B. Mortazavi}. The C$_3$N nanoribbons may be obtained by cutting a monolayer two-dimensional C$_3$N nanosheet along with the direction of armchair edge or zigzag edge, i.e., armchair edge and zigzag edge nanoribbons. According to the arrangement of C and N atoms in the C$_3$N nanosheet, the C$_3$N nanoribbons exist two different edge atomic configurations, named as CC-edge and CN-edge. Here the CC-edge signifies the outmost atoms are composed of four C atoms while the CN-edge signifies the outmost atoms include two C and two N atoms. These two types of edge atomic configurations would generate three different AC$_3$NNRs, in which two of them have the same edge configuration at both edges. Namely, the one having CC-edge at both edges is defined as CCA in Fig. 1(a) and the one possessing two CN-edges are called as CNA in Fig. 1(b). As shown in Fig. 1(c), the third type is the AC$_3$NNRs with combined edge configurations of CC-edge and CN-edge, which means one edge is composed of only C atoms and the other edge is composed of both C and N atoms, named as CCCNA. Likewise, a C$_3$N nanosheet can also be cut into three kinds of ZC$_3$NNRs \cite{Y. Ren}. To weaken the interaction between the two edges of nanoribbons, the width of constructed C$_3$N nanoribbons are both about 10 {\AA}. After adequate structure optimization, the structure of these C$_3$N nanoribbons are well maintained as planes and no significant space warp, which manifest that these ribbons would have existed as stable plane structures like graphene.

In this paper, we mainly explore the AC$_3$NNRs. There are two reasons for without considering the ZC$_3$NNRs. First, through the calculated total energy and thermal stability of the ribbons in the previous calculation, we found that AC$_3$NNRs show better stability than ZC$_3$NNRs \cite{Y. Ren}. Second, we mainly focus on rectification and the establishment of Schottky heterostructure. The band structures show that pristine ZC$_3$NNRs are ferromagnetic metal, while the H-passivated ZC$_3$NNRs have only one type of nonmagnetic semiconductor \cite{Congxin Xia,Y. Ren}. Furthermore, the nature of AC$_3$NNRs have not been revealed yet. Therefore, we construct the heterojunction models using the pristine and H-passivated AC$_3$NNRs. Since the crystal structures of pristine and H-passivated AC$_3$NNRs are identical, there is no lattice mismatch at the interface. Further, the unpassivated AC$_3$NNRs are metal and the H-passivated AC$_3$NNRs are semiconductor. So the in-plane contact structure leads to the formation of a metal-semiconductor heterojunction, which results in rectifying current-voltage characteristics. The optimized atomic of the heterojunctions are obtained using the following tactics: first the optimized pristine AC$_3$NNRs and H-passivated AC$_3$NNRs are stitched together, then the positions of all atoms are relaxed except for the atoms of the lead and the lead extension, finally the heterojunctions are optimized until the convergence criterion of force is reached. After the structure optimization, the two components of the heterojunction are contacted seamlessly.

The geometric optimization, the electronic structure and the subsequent electronic transport calculations of our constructed heterojunctions are implemented by first-principles software package Atomistix ToolKit, which is based on DFT combing the non-equilibrium Green's function (NEGF) technique \cite{M. Brandbyge,J. Taylor}. The Perdew-Burke-Ernzerhof (PBE) formulation of the generalized gradient approximation (GGA) is used as the exchange-correlation function \cite{G. Kresse,Soler J. M}. Double-zeta plus polarization (DZP) basis set is used for all elements including carbon, hydrogen and nitrogen atoms. The Brillouin zone of the electrode is sampled as a Monkhorst-Pack grid using 1$\times$1$\times$100 $k$-points and adopt 150 Ry cutoff energy to achieve the balance between efficiency and calculation accuracy \cite{Monkhorst H. J}. The energy convergence criterion of self-consistent calculations is 10$^{-6}$ eV and the geometries are optimized until all residual forces on each atom are smaller than 0.01 eV{{\AA}}$^{-1}$ and the electrode temperature is set to be 300 K. In our calculations, at least 15 {\AA} vacuum buffer space is used to eliminate interaction between the model. When a bias voltage is applied, the current is calculated by using the Landauer-B$\rm\ddot{u}$ttiker formula \cite{H. M. Pastawski}
\begin{equation}
I(V)=\frac{2e}{h}\int T(E,V)[f(E-\mu_L)-f(E-\mu_R)],
\end{equation}
where $e$ is the electron charge, $h$ is the Planck constant, $\mu$$_{L/R}$ (=$E_F\mp eV$/2) is the chemical potential of the left/right electrode, \emph{f}(\emph{E}-$\mu$$_{L/R}$) is the Fermi-Dirac distribution functions of the left/right electrode, and \emph{V} is the bias window. The transmission coefficient can be obtained by using Green's function
\begin{equation}
T(E,V)=\text{Tr}[\Gamma_L(E,V)G^R(E,V)\Gamma_R(E,V)G^A(E,V)],
\end{equation}
where \emph{G}$^{A(R)}$(\emph{E},\emph{V}) is the advanced (retarded) Green's function and $\Gamma$$_{L(R)}$ is the coupling matrix to the left (right) electrode.

\section{Results and Discussions}

\subsection{The electronic structure of pristine AC$_3$NNRs}

\begin{figure}
\center
\includegraphics[bb=118 91 501 552, width=3.3 in]{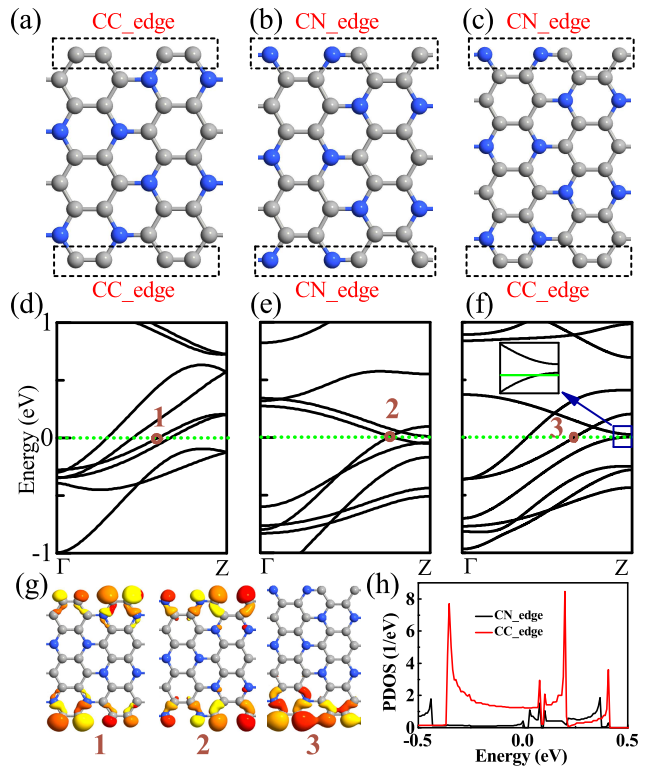}
\renewcommand{\figurename}{Fig}
\caption{(Color online) (a)-(c) The super cells of AC$_3$NNRs with three possible different combinations of edge atomic species and configurations, denoted as CCA, CNA and CCCNA, respectively. The gray and blue balls stand for C atom and N atom,respectively. The energy band structures of three possible AC$_3$NNRs, with (d)-(f) corresponding to CCA, CNA and CCCNA, respectively. (g) The spatial distributions of Bloch states corresponding to states 1-3 pointed by the brown points in (d)-(f), where the isovalues are fixed at $|e|$/{\AA}$^3$. (h) The projected density of state (PDOS) as a function of energy for the CCCNA nanoribbon.}
\end{figure}

\begin{figure}
\center
\includegraphics[bb=120 72 483 507, width=3.3 in]{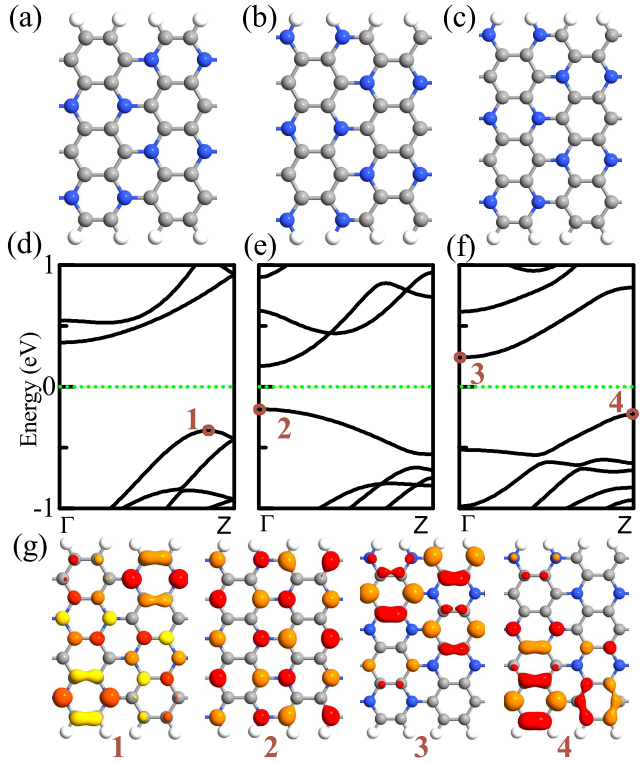}
\renewcommand{\figurename}{Fig.}
\caption{(Color online) (a)-(c) The super cells of AC$_3$NNRs (corresponding to Figs. 1 (a-c)) after hydrogen passivation named H-CCA, H-CNA and H-CCCNA, where the gray, blue and white balls stand for C, N and H atoms, respectively. The band structures of (d) H-CCA, (e) H-CNA and (f) H-CCCNA. (g) The spatial distributions of Bloch states corresponding to states 1-4 pointed by the brown points in (d)-(f), where the isovalues are fixed at $|e|$/{\AA}$^3$.}
\end{figure}

To study the effect of nanoribbons edge configuration on the electronic structures of AC$_3$NNRs, the band structures of CCA, CNA and CCCNA are calculated, respectively. The detailed results are shown in Figs. 1(d-f). As is seen in Fig. 1(d), the electronic band structure of CCA along a high-symmetry path in the Brillouin zone is shown that have four bands crossing the Fermi level, so the CCA ribbon is metal. Additionally, our results also show that the CNA and CCCNA are metal, because there are two bands and three bands across the Fermi level in the band structures of CNA and CCCNA as shown in Figs. 1(e) and 1(f), respectively. So the pristine AC$_3$NNRs are metal, which is in good agreement with other theoretical calculations \cite{Y. Ren}. Interestingly, we find that these ribbon's edge configuration and band structures have a law with the CN-edge changes. The number of bands crossing the Fermi level gradually decreases from 4 to 2 as the number of CN-edge in ribbon increases from 0 to 2. And the H-passivated ZC$_3$NNRs have the same electronic band \cite{Congxin Xia}. It is well known that the electronic states near the Fermi level play an important role in the electron property. To further understand why the bands across the Fermi level are decreasing, we give the spatial Bloch state distributions of AC$_3$NNRs at the band structure near the Fermi level.

Further, as can be seen from Fig. 1(g), the contours of Bloch states of pristine AC$_3$NNRs are dominated by the atoms near the edges. The color in the picture only represents the phase, and the size represents how much the electron contributes at the Fermi level \cite{Cao1,Y. Ren,Q. Y. Wu}. Further, we find that the electronic density of edge C atoms is much larger than that of the edge N atoms for each ribbon. This is very similar to the situation in graphene nanoribbon, in which the energy bands near $E_F$ are mainly contributed by the edge C atoms, and the contribution to the electron orbital is mainly due to the $p$-orbital near the Fermi level. The graphene-like AC$_3$NNRs here also show a very strong edge state effect. In addition, we also calculate the projected density of state (PDOS) of CCCNA as depicted in Fig. 1(h), calculated the contribution of the CC-edge and the CN-edge to the density of state (DOS) near the Fermi level. PDOS analysis for CCCNA shows that the CN-edge have little contribution to the DOS near the Fermi level. This similar phenomenon also occurs in B and N-doped zigzag graphene nanoribbons \cite{S. Dutta}. It is found that electrons flow preferentially from C to B, but the transfer of charge from C to N is hindered by the existence of large Coulomb repulsion at the N position. As a result, the N atoms and the connected C atoms contribute very little to DOS near the Fermi level. So in our nanoribbon structures, it is the large Coulomb repulsion on N atoms that hinders the electron transfer from the nearest C atoms, which causes the CN-edge to contribute little to the DOS near the Fermi level, and further causes the state of the band crossing the Fermi level to be affected, thereby reducing its number.

\subsection{The electronic structure of H-passivated AC$_3$NNRs}
Considering the effect of edge passivation on the nanoribbons, we tested the passivation of CCA, CNA and CCCNA ribbon edge by H atoms. As shown in Figs. 2(a-c), the three kinds of AC$_3$NNRs after edge terminate in Hydrogen atoms were named H-CCA, H-CNA and H-CCCNA, respectively. Through the calculated edge energy of the ribbons of bare edges and hydrogen passivation, we found that all the AC$_3$NNRs edges passivated by H atoms can enhance the stability of the ribbons. Not only that, but hydrogen passivation can also tune the bandgap of the ribbons. The corresponding band structures after hydrogen passivation are illustrated in Figs. 2(d-f). From the CBM and VBM at $\Gamma$ point and along the $\Gamma$-Z line shown in Fig. 2(d), we know the H-CCA is an indirect band gap semiconductor with a gap value of 0.7 eV. Interestingly, the H-CNA structure possesses the direct bandgap with a gap value of 0.35 eV in Fig. 2(e). So H-passivation achieves the indirect-direct gap transition in nanostructure of C$_3$N. In addition, the H-CCCNA structure is an indirect gap semiconductor with a gap value of 0.46 eV in Fig. 2(f). Therefore, for AC$_3$NNRs, H-passivation can cause the bands acrossing the Fermi level to disappear and thus induce the transition from metal to semiconductor. In other words, the unpassivated AC$_3$NNRs are metal while the H-passivated AC$_3$NNRs are semiconductor. The reason for the edge states vanishing and the insulating band appearing is the electrons from H-atom bonds with unpaired electrons of outermost edge atoms, and the edge state is full-filled and buried deeply in the valence band. The zigzag phosphorene nanoribbon also has the similar electronic structure \cite{Fan}.

\begin{figure}[b]
\center
\includegraphics[bb=133 129 580 503, width=3.3 in]{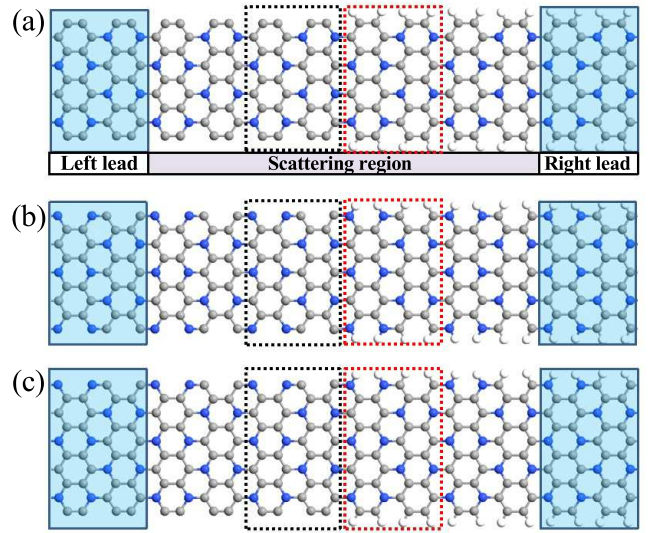}
\renewcommand{\figurename}{Fig.}
\caption{(Color online) Schematic illustrations of C$_3$N heterojunctions constructed of armchair C$_3$N structures with different edges. (a) CCA + H-CCA (M1); (b) CNA + H-CNA  (M2); (c) CCCNA + H-CCCNA (M3).}
\end{figure}

To further investigate the effect of H-passivation on the electronic property of AC$_3$NNRs, we select a few points on the energy band diagram and plotted its electron density. Hence states 1-4 of three passivated AC$_3$NNRs spatial distribution of Bloch state are shown in Fig. 2(g). Observing the respective Bloch functions, it is first observed that the edge states disappear in the H-passivated AC$_3$NNRs. Secondly, the contours of Bloch state indicate that both CBM and VBM are contributed by the hybridized \emph{s-p} states of the C and N atoms of the AC$_3$NNRs. The outline of the Bloch state shows that the VBM of H-CCA (state 1) and the VBM of H-CNA (state 2) are both contributed by central atoms in the ribbons. The Bloch state of their CBM is not given in this paper because its contour is similar to the Bloch state of VBM. Furthermore, the states of electron density are distributed almost symmetrically throughout the structure. This is because of the inversion symmetry of the CCA and the mirror symmetry of the CNA ribbon. On the contrary, for H-CCCNA, the contours of the Bloch state indicate that both CBM (state 3) and VBM (state 4) are contributed by edge atoms and partial central atoms in the ribbon, and that the states are gradually localized towards the edge of the ribbon. This is caused by the asymmetry structure for the H-CCCNA ribbon. Namely, the H-CCCNA transport channels for carrier transport under a low bias are different from the other two structures.

\subsection{Transport property of AC$_3$NNR-based junctions}
Based on the knowledge that the pristine and H-passivated AC$_3$NNRs are metal and semiconductor, we build the heterojunction models of Schottky junctions in Fig. 3, where the left lead and right lead are the unit cell of pristine and H-passivated AC$_3$NNRs, respectively. The central scattering regions are constructed by two unit cells of metallic AC$_3$NNRs and corresponding to two unit cells of semiconducting H-passivated AC$_3$NNRs. The interface is the border between the black and red dashed boxes. For the sake of the simplicity, we named M1 of the heterojunction with CCA on the left and H-CCA on the right in Fig. 3(a). Likewise, Figs. 3(b-c) show the CNA/H-CNA heterojunction (M2) and CCCNA/H-CCCNA heterojunction (M3), respectively.

\begin{figure}
\center
\includegraphics[bb=72 35 726 495, width=3.3 in]{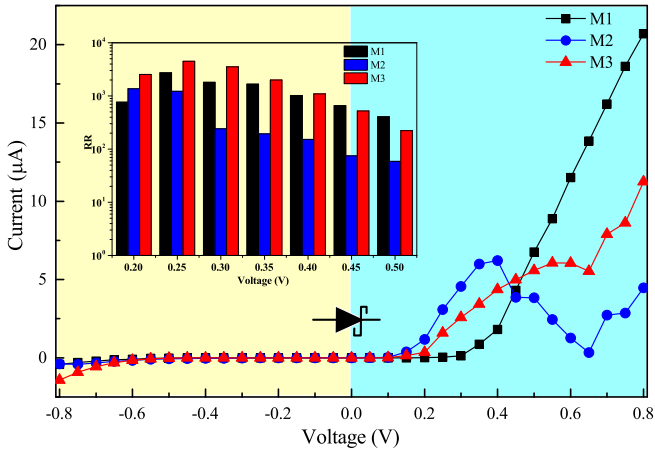}
\renewcommand{\figurename}{Fig.}
\caption{(Color online) The current-voltage characteristics of three Schottky junctions in a bias region of -0.8 to 0.8 V. The inset figure is corresponding rectification ratio (RR) of the planar Schottky contact structures in the bias region from 0.2 V to 0.4 V.}
\end{figure}

The current-voltage (\emph{I-V}) characteristics and rectification ratio of planar metal-semiconductor junctions are illustrated clearly in Fig. 4. Interestingly, all heterojunction models show forward-conducting and reverse-blocking rectifier diode characteristics. In other words, all currents increase rapidly under positive bias, yet they are still almost zero under negative bias. But the difference is that when the M2 junction bias is 0.4 V-0.65 V, the current decreases as the voltage increases, showing a negative differential resistance (NDR) effect. To precisely describe the rectification behaviors, we introduced the rectification ratio. It is defined as RR= I$_+$/I$_-$, where the I$_+$ and I$_-$ are the currents at positive and negative bias for the same bias magnitude. The rectification ratios of three junctions under different bias are exhibited in the inset figure of Fig. 4, showing that the junction of M3 has larger rectification ratio under the 0.2 V-0.4 V bias. And it's important to note that the RR depends on the ratio between positive and negative voltages by definition. Although the current of M2 is greater than that of M3 under positive bias voltage, the former is much smaller than the latter under negative bias voltage, so the RR of M3 is greater than that of M2 in Fig. 4.

\begin{figure}
\center
\includegraphics[bb=2 3 511 349, width=3.4 in]{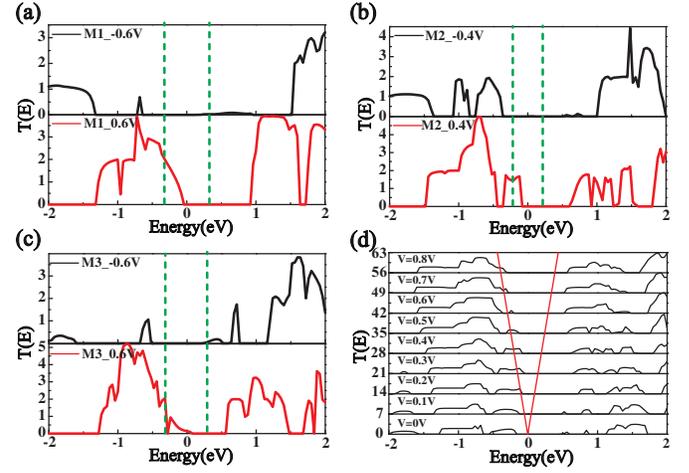}
\renewcommand{\figurename}{Fig.}
\caption{(Color online) The transmission spectra of (a) M1 under bias of -0.6 and 0.6 V; (b) M2 under bias of 0.4 and -0.4 V; (c) The M3 junction under bias of 0.6 and -0.6 V.  (d) M2 under bias of 0 - 0.8 V, the bias increases with a step of 0.1 V from bottom to up and the curve is shift up by 7 units with gradually increasing the bias. The area between the green  dashed lines and solid red lines is the bias window. Zero energy is the Fermi level.}
\end{figure}

To understand the rectification behavior, transmission spectra of M1, M2 and M3 at several representative bias voltages are shown in Figs. 5(a-c). The green dashed line represent the bias window. According to the Landauer-B$\rm\ddot{u}$ttiker formula \cite{H. M. Pastawski}, the value of the current passing through the central scattering region depends on the integrated area of the transmission probability within the bias window under consideration. For the M1 junction, as shown in Fig. 5(a), when a negative bias of -0.6 V is applied, there is no transmission peak in the bias window, therefore the electrons in this system cannot be transmitted, so the current shown in the \emph{I-V} curve is almost zero. When a positive bias of 0.6 V is applied, there is a transmission peak in the bias window, so the integral area of the transmission probability in the bias window is not zero, therefore, the current reflected on the \emph{I-V} curve has a certain value.

\begin{figure}
\center
\includegraphics[bb=36 81 759 455, width=3.3 in]{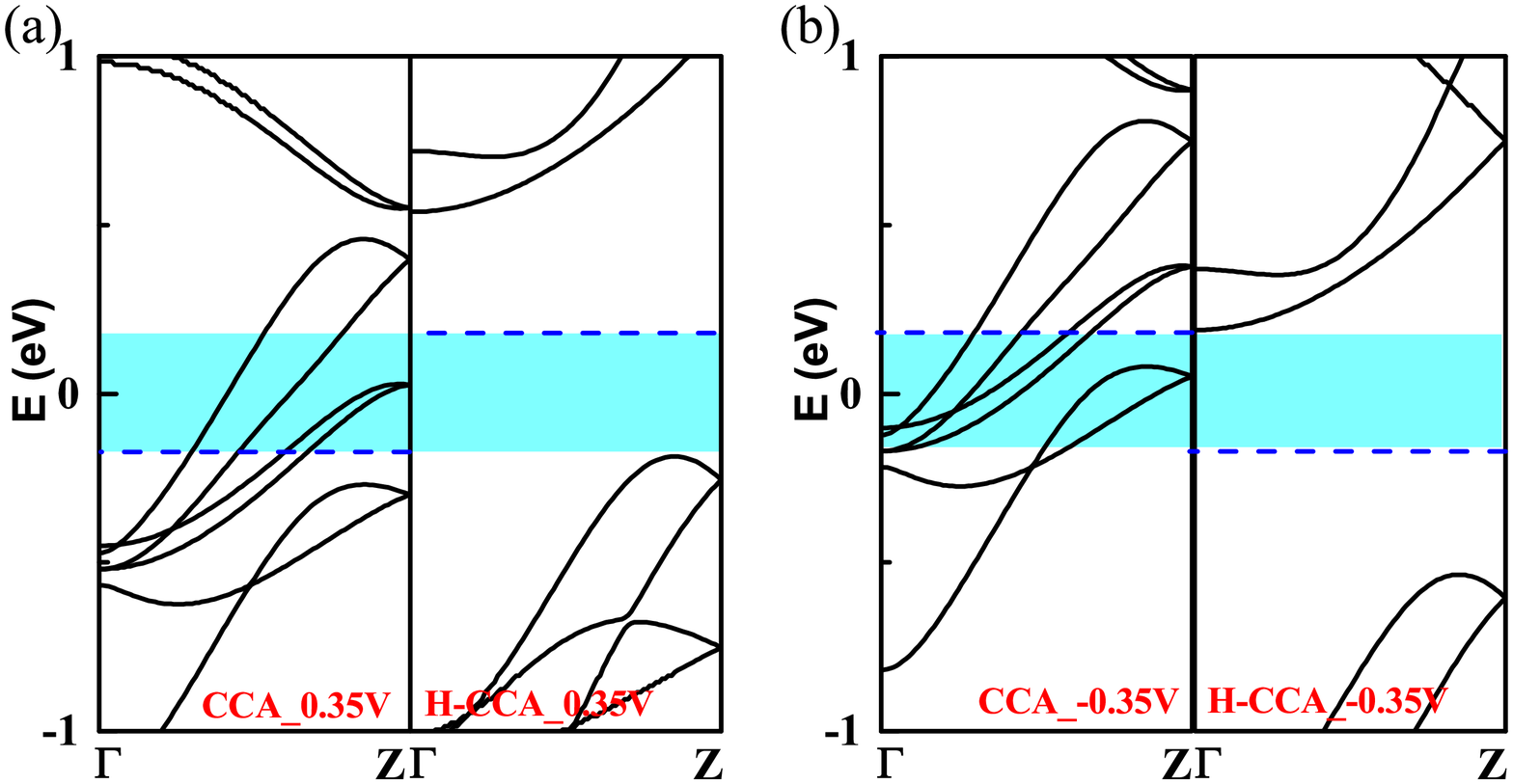}
\renewcommand{\figurename}{Fig.}
\caption{(Color online) Band structures around the Fermi level of the left and right electrodes of M1 junction (a) under forward bias of 0.35 V; (b) under reverse bias of -0.35 V.  The blue dotted line is the Fermi level. The light blue rectangular area is the energy window.}
\end{figure}

The \emph{I-V} curves become asymmetrical with those at positive and negative bias, indicating rectification behavior. In Figs. 5(b-c), the rectification mechanism of M2 and M3 junctions is the same as that of M1 junction. To precisely understand the mechanism of the NDR behavior appearing in M2 structure, we draw the transmission spectrum in Fig. 5(d). The red solid lines represents the bias window. It is clear that the integral area in the bias window first increased, then decreased, and increased again as the bias voltage increased, which is consistent with the \emph{I-V} curve of M2 in Fig. 4. Thus, the conductivity of this junction was high at 0.3 V-0.4 V biases, whereas it is intensively suppressed at 0.5 V-0.6 V biases suggesting the NDR behavior emerged. It is well known that, the energy band structure of the material determines the property of electron transport, which will be discussed in detail in the next section.

\begin{figure*}
\center
\includegraphics[bb=0 0 526 146, width=6.6 in]{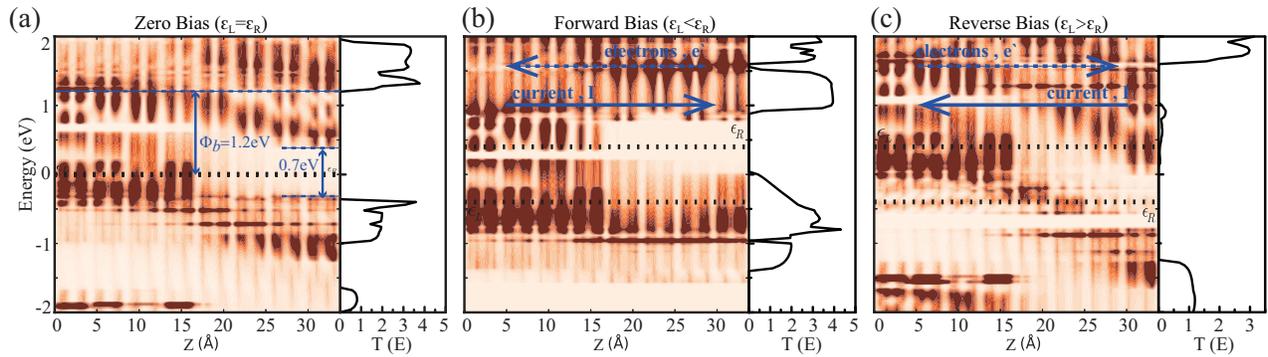}
\renewcommand{\figurename}{Fig}
\caption{(Color online) The projected local density of states (PLDOS) plot (left panels) and corresponding transmission spectrum (right panels) of the M1 junction (a) at 0 bias, (b) under forward bias of 0.8 V, and (c) under reverse bias of -0.8 V. The abscissa represents the Cartesian coordinate of the central area along the Z direction. The horizontal black dashed represents the Fermi level. The shade of color indicates the DOS size.}
\end{figure*}

\begin{figure*}
\center
\includegraphics[bb=16 112 779 510, width=6.6 in]{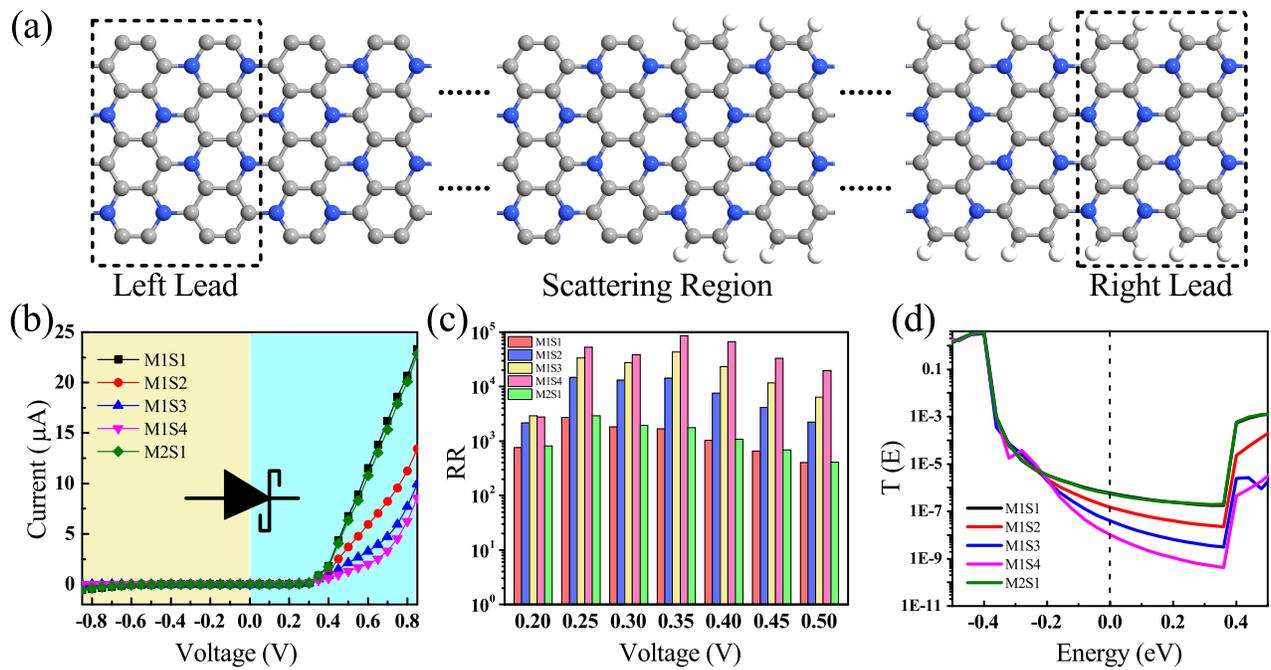}
\renewcommand{\figurename}{Fig.}
\caption{(Color online) The schematic structures (a), \emph{I-V} curves (b), rectification ratio (c) and transmission spectrum at zero bias (d) of Schottky junctions based on CCA and H-CCA ribbon with different channel length.}
\end{figure*}

Next, choosing M1 as the representative junction to analyze, the left and right electrodes are CCA and H-CCA, respectively. If we apply a bias to the electrodes, there will be a potential difference between them, and their band structures will move downward or upward relatively. We apply a forward bias of 0.35 V in Fig. 6(a), and an energy window (light blue shaded area) appears due to Fermi level migration. Band structures of CCA and H-CCA without applied voltage are shown in Figs. 1(d) and 2(d), respectively. Compared with Fig. 6(a), it is found that the Fermi level of the left electrode will move downward when the positive voltage is applied, while the Fermi level of the right electrode will move upward. In this energy window, only the left electrode has several bands while the right electrode has a band gap between the VBM and the energy window, so electrons cannot jump from the right electrode to the left electrode, resulting in electron transmission is prohibited. As the forward bias continues to increase, the valence band of the right electrode will enter the energy window, causing electrons can jump between the two electrodes, and the junction opens. This is why the positive cut-off bias of M1 in Fig. 4 is 0.35 V. Fundamentally speaking, the forward cut-off bias depends on the absolute value of the right electrode VBM, and the larger the value, the greater the cut-off bias. This value is about -0.18 eV for H-CNA, and about -0.22 eV for H-CCCNA, so the forward cut-off biases for their corresponding junctions are 0.18 V and 0.22 V, respectively, as shown in Fig. 4.

Further, after applying a reverse bias of -0.35 V, the band structures of the left and right electrodes are shown in Fig. 6(b). Compared with a positive voltage, applying a negative voltage causes the Fermi level of the left electrode to move up and the Fermi level of the right electrode to move down. Like the forward bias mechanism, due to the band gap between the CBM of the right electrode and the energy window, electrons cannot jump within the energy window. The difference is that when the bias voltage increases, the right electrode conduction band can enter the energy window and electrons can jump, but the reverse current in Fig. 4 is still zero. We find that this situation also exists in the M3 junction, so there is no jumping restriction caused by symmetry. In this case, we suspect that there is a potential barrier prevents electrons from being transmitted. Therefore, we calculate the projected local density of states (PLDOS) shown in Fig. 7 to examine the interface barriers under different bias voltages.

Moreover, as shown in Fig. 7(a), it is observed that the left part is metallic, while the right part is a semiconductor with a band gap of about 0.7 eV which is consistent with the analysis of Fig. 2(d). There are several forbidden bands appear on either side of the heterojunction under the near Fermi level. The energy region of forbidden bands are from 0.65 to 0.75 eV in the CCA and from -0.35 to 0.65 eV in the H-CCA. These interleaved forbidden bands account for the zero transmission in the energy regimes from -0.35 to 0.75 eV. However, the density of states exists on both sides of the heterojunction in the energy range of 0.75 to 1.2 eV, and the transmission also disappears at this time. The transmission disappeared in this energy region may be caused by other reasons. As we all know, when two materials in a metal-semiconductor junction are in contact, the carrier diffusion flow must make the Fermi energy levels on both sides of the contact surface equal to reach an equilibrium state. Therefore, as shown in Fig. 7(a), the energy band in the semiconductor will be bent downwards due to the built-in electric field after contact. In this way, an electron potential barrier is formed on the contact surface, called the Schottky barrier. The Schottky barrier height ($\Phi$$_b$) is 1.2 eV, as seen in Fig. 7(a). The transport forbidden in the energy region from 0.75 to 1.2 eV is explained. In Fig. 7(b), when a forward bias of 0.8 V is applied, the addition of an external electric field raises the Fermi level on the semiconductor side. So the chemical potential of the right electrode is higher than the energy of the left electrode, the electrons flow from the right to the left electrode, and current flows from the left to the right electrode. Because the Fermi level on the right side rises,the CBM and VBM near the Fermi level are flattened or bent upward \cite{W. X. Zhang}.

Also, we can find that the electrons do not need to overcome any potential barriers on the interface during the flow, so the M1 junction has a larger current under high forward bias. On the contrary, after applying a reverse bias of -0.8 V, the Fermi level on the metal side rises and the Fermi level on the semiconductor side falls as shown in Fig. 7(c). So the chemical potential of the left electrode is higher in energy than the chemical potential of the right electrode, and electrons flow from the left to the right electrode. However, although the left Fermi level moves upward under the reverse bias, its energy is still lower than the energy of the CBM on the right side at the interface, so a potential barrier appears at the interface \cite{Z. X. Yang,W. X. Zhang}. In the process of flowing from the left electrode to the right electrode, electrons must overcome this potential barrier, so it generate a very low current. Therefore the M1 junction is off at reverse bias. Fundamentally, the potential barrier is still due to the semiconductor nature with the wide bandgap in the right part of heterojunction.

\subsection{Channel length effect on transport in junctions}
In order to explore the effect of channel length on transport property, as shown in Fig. 8(a), we chose CCA and H-CCA ribbon to construct lateral heterojunction devices. The central scattering region (except for electrode extension) which contains one unit cell of metallic CCA and one unit cell of semiconducting H-CCA is named M1S1. After the CCA increases to two unit cells, the structure is named as M2S1. When the H-CCA extends to two unit cells, three unit cells, and four unit cells, the structures are named as M1S2, M1S3 and M1S4, respectively. The \emph{I-V} curve of five Schottky contact structures are presented in Fig. 8(b). When the applied positive bias exceeds 0.35 V, for each Schottky junction, the current increases rapidly with increasing the positive bias. This is consistent with the forward cut-off bias of 0.35 V calculated in the previous section. For M2S1, the length of the CCA ribbon in scattering region is twice that of the H-CCA ribbon, and the \emph{I-V} curve nearly overlaps with that of M1S1. By expanding the length of H-CCA ribbon in scattering region, the currents of the lateral heterostructure decrease accordingly. The reason can be given as follows. The transmission spectrum near the Fermi level at zero bias can evaluate the current of junctions under low bias. From Fig. 8(d), the M1S2, M1S3 and M1S4 transmission coefficient near the Fermi level is smaller, which induces the decrease of current. Interesting, although the current value decreases under both positive and negative bias, the rectification ratio [Fig. 8(c)] can be significantly increased. In Fig. 8(c), we can see that the junctions of expanding H-CCA ribbon have larger rectification ratio under the same bias.

Furthermore, the longer the extended length, the greater the rectification ratio \cite{Fan}. The rectification ratios of M1S2, M1S3 and M1S4 junctions possess largest values at 0.35 V, and the largest value of M1S4 junction arrives to the magnitude of 10$^5$. Discussed above demonstrate increasing the semiconductor length of Schottky junctions based on CCA and H-CCA ribbon will block the transmission of electrons at the interface and reduce of the transmission coefficients. This is the reason why the current will decrease, but current value is also have an extreme, which will not always decrease, and this can be clearly seen from the \emph{I-V} curve. But expanding the metal length of Schottky junctions causes transport property almost unchanged.

\section{Conclusions}
In summary, we have investigated the band structures of three pristine and H-passivated AC$_3$NNRs and transport properties of the planar heterostructures by using density-functional theory in combination with the non-equilibrium Green's function. The study found that the edge morphology of AC$_3$NNRs resulted in different electronic structures and electron transport property. The pristine AC$_3$NNRs are metal, while the H-passivated AC$_3$NNRs reveal semiconductor. Then, we use three pristine AC$_3$NNRs and their corresponding H-passivated nanoribbon to fabricate lateral planar Schottky junctions. We found that these three types of heterojunctions all show forward-conducting and reverse-blocking rectifier diode behavior, and the M2 junction shows the NDR effect at 0.4 V-0.65 V. This is mainly caused by the band structure of the right electrode and the interface potential barrier. In addition, by expanding the length of semiconductor in scattering region, we find that the longer heterojunction channel possesses larger rectification ratio, even the rectification ratio up to 10$^5$. Our research results are expected to further enhance the application value of AC$_3$NNRs in future electronic devices.

\begin{acknowledgments}
This work was supported by the National Natural Science Foundation of China (Grant Nos. 61801520 and 11774085), and the Scientific Research Project of Hunan Provincial Education Department (Grant No. 20B144).
\end{acknowledgments}


\begin{references}
\bibitem{Novoselov1} K. S. Novoselov, A. K. Geim, S. V. Morozov, D. Jiang, Y. Zhang, S. V. Dubonos, I. V. Grigorieva and A. A. Firsov, Electric Field Effect in Atomically Thin Carbon Films,
\textcolor{blue}{Science {\bf 306}, 666 (2004)}.
\bibitem{Novoselov2} K. S. Novoselov, A. K. Geim, S. V. Morozov, D. Jiang, M. I. Katsnelson, I. V. Grigorieva, S. V. Dubonos and A. A. Firsov, Two-dimensional gas of massless Dirac fermions in graphene,
\textcolor{blue} {Nature {\bf 438}, 197 (2005)}.
\bibitem{Castro Neto} A. H. Castro Neto, F. Guinea, N. M. R. Peres, K. S. Novoselov, and A. K. Geim, The electronic properties of graphene,
\textcolor{blue} { Rev. Mod. Phys. {\bf 81}, 109 (2009).}
\bibitem{Strano} Q. H. Wang, K. Kalantar-Zadeh, A. Kis, J. N. Coleman, and M. S. Strano, Electronics and optoelectronics of two-dimensional transition metal dichalcogenides,
\textcolor {blue} { Nat. Nanotechnol. {\bf7}, 699 (2012).}
\bibitem{Chhowalla} M. Chhowalla, H. S. Shin, G. Eda, L. J. Li, K. P. Loh, and H. Zhang, The chemistry of two-dimensional layered transition metal dichalcogenide nanosheets,
\textcolor{blue} { Nat. Chem. {\bf5}, 263 (2013).}
\bibitem{Li} L. K. Li, Y. J. Yu, G. J. Ye, Q. Q. Ge, X. D. Ou, H. Wu, D. L. Feng, X. H. Chen and Y. B. Zhang, Black phosphorus field-effect transistors,
\textcolor{blue} { Nat. Nanotechnol. {\bf9}, 372 (2014).}
\bibitem{Liu} Han Liu, Adam T. Neal, Zhen Zhu, Zhe Luo, Xianfan Xu, David Tomanek, and Peide D. Ye, Phosphorene: An Unexplored 2D Semiconductor with a High Hole Mobility,
\textcolor{blue} {ACS Nano. {\bf8}, 4033 (2014).}
\bibitem{Qiao} Jingsi Qiao, Xianghua Kong, Zhi-Xin Hu, Feng Yang and Wei Ji, High-mobility transport anisotropy and linear dichroism in few-layer black phosphorus,
\textcolor{blue} {Nat. Commun. {\bf5}, 4475 (2014).}
\bibitem{Xia} Fengnian Xia, Han Wang, and Yichen Jia, Rediscovering black phosphorus as an anisotropic layered material for optoelectronics and electronics,
\textcolor{blue} {Nat. Commun. {\bf5}, 4458 (2014).}
\bibitem{Baugher} B. W. H. Baugher, H. O. H. Churchill, Y. F. Yang, P. Jarillo-Herrero, Optoelectronic devices based on electrically tunable p-n diodes in a monolayer dichalcogenide,
\textcolor{blue} { Nat. Nanotechnol. {\bf9}, 262 (2014).}
\bibitem{Cao1} L. Cao, X. Li, M. Zuo, C. Jia, W. Liao, M. Long and G. Zhou, Perfect negative differential resistance, spin-filter and spin-rectification transport behaviors in zigzag-edged $\delta$-graphyne nanoribbon-based magnetic devices,
\textcolor{blue} { J. Magn. Magn. Mater. {\bf485}, 136 (2019).}
\bibitem{Habib} K. M. M. Habib and R. K. Lake, Current modulation by voltage control of the quantum phase in crossed graphene nanoribbons,
\textcolor{blue} { Phys. Rev. B {\bf86}, 045418 (2012).}
\bibitem{Wang} Z. F. Wang, Q. Li, Q. W. Shi, X. Wang, J. Yang, J. G. Hou and J. Chen, Chiral selective tunneling induced negative differential resistance in zigzag graphene nanoribbon: A theoretical study,
\textcolor{blue} { Appl. Phys. Lett. {\bf92}, 133114 (2008).}
\bibitem{H. Ren} H. Ren, Q. X. Li, Y. Luo and J. Yang, Graphene nanoribbon as a negative differential resistance device,
\textcolor{blue} { Appl. Phys. Lett. {\bf94}, 173110 (2009).}
\bibitem{Wan1} H. Q. Wan, B. H. Zhou, X. W. Chen, C. Q. Sun and G. H. Zhou, Switching, Dual Spin-Filtering Effects, and Negative Differential Resistance in a Carbon-Based Molecular Device,
\textcolor{blue} { J. Phys. Chem. C {\bf116}, 2570 (2012).}
\bibitem{Moon} J. S. Moon, H.-C. Seo, F. Stratan, M. Antcliffe, A. Schmitz, R. S. Ross, A. A. Kiselev, V. D. Wheeler, L. O. Nyakiti, D. K. Gaskill, K.-M. Lee, and P. M. Asbeck, Lateral Graphene Heterostructure Field-Effect Transistor,
\textcolor{blue} { IEEE Electron Dev. Lett. {\bf34}, 1190 (2013).}
\bibitem{Karpan} V. M. Karpan, G. Giovannetti, P. A. Khomyakov, M. Talanana, A. A. Starikov, M. Zwierzycki, J. van den Brink, G. Brocks, and P. J. Kelly, Graphite and Graphene as Perfect Spin Filters,
\textcolor{blue} { Phys. Rev. Lett. {\bf99}, 176602 (2007).}
\bibitem{Wan2} H. Q. Wan, B. H. Zhou, W. H. Liao, and G. H. Zhou, Spin-filtering and rectification effects in a Z-shaped boron nitride nanoribbon junction,
\textcolor{blue} { J. Chem. Phys. {\bf138}, 034705  (2013).}
\bibitem{Cao2} L. M. Cao, X. B. Li, C. X. Jia, G. Liu, Z. R. Liu, G. H. Zhou, Spin-charge transport properties for graphene/graphyne zigzag-edged nanoribbon heterojunctions: A first-principles study,
\textcolor{blue} { Carbon {\bf127}, 519 (2018).}
\bibitem{Jiang} X. W. Jiang, and S. S. Li, Performance limits of tunnel transistors based on mono-layer transition-metal dichalcogenides,
\textcolor{blue} { Appl. Phys. Lett. {\bf104}, 193510 (2014).}
\bibitem{Cao3} L. M. Cao, G. H. Zhou, Q. Y. Wu, Shengyuan A. Yang, H. Y. Yang, Yee Sin Ang, and L.K. Ang, Electrical Contact between an Ultrathin Topological Dirac Semimetal and a Two-Dimensional Material,
\textcolor{blue} { Phys. Rev. Appl. {\bf13}, 054030 (2020).}
\bibitem{Z. X. Yang} Z. X. Yang, J. L. Pan, Q. Liu, N. N. Wu, M. L. Hu and F. P. Ouyang, Electronic structures and transport properties of a MoS$_2$-NbS$_2$ nanoribbon lateral heterostructure,
\textcolor{blue} { Phys. Chem. Chem. Phys. {\bf19}, 1303 (2017).}
\bibitem{Joshua} Joshua O Island, Gary A Steele, Herre S J van der Zant and Andres Castellanos-Gomez, Environmental instability of few-layer black phosphorus,
\textcolor{blue} { Sci. Rep. {\bf5}, 8989 (2015).}
\bibitem{Desai} S. B. Desai, S. R. Madhvapathy, A. B. Sachid, J. P. Llinas, Q. Wang, G. H. Ahn, G. Pitner, M. J. Kim, J. Bokor, C. Hu, H.-S. Philip Wong and Ali Javey, MoS$_2$ transistors with 1-nanometer gate lengths,
\textcolor{blue} { Science {\bf354}, 99 (2016).}
\bibitem{Yu} Z. H. Yu, Z.-Y. Ong, S. L. Li, J.-B. Xu, G. Zhang, Y.-W. Zhang, Y. Shi, and X. R. Wang, Analyzing the Carrier Mobility in Transition-Metal Dichalcogenide MoS$_2$ Field-Effect Transistors,
\textcolor{blue} { Adv. Funct. Mater., 1604093 (2017).}
\bibitem{Yang} S. Yang, W. Li, C. Ye, G. Wang, H. Tian, C. Zhu, P. He,G. Ding, X. Xie, Y. Liu, Y. Lifshitz, S.-T. Lee, Z. Kang and M. Jiang, C$_3$N-A 2D Crystalline, Hole-Free, Tunable-Narrow-Bandgap Semiconductor with Ferromagnetic Properties,
\textcolor{blue} { Adv. Mater., 1605625 (2017).}
\bibitem{A. Bafekry} A. Bafekry, C. Stampfl, and S. Farjami Shayesteh, A first-principles study of C$_3$N nanostructures: Control and engineering of the electronic and magnetic properties of nanosheets, tubes and ribbons, \textcolor{blue} { ChemPhysChem {\bf21}, 164 (2020).}
\bibitem{B. Mortazavi} B. Mortazavi, Ultra high stiffness and thermal conductivity of graphene like C$_3$N,
\textcolor{blue} { Carbon {\bf118}, 25 (2017).}
\bibitem{Y. Ren} Y. Ren, F. Cheng, X. Y.  Zhou, K. Chang, G. H. Zhou, Tunable mechanical, electronic and magnetic properties of monolayer C$_3$N nanoribbons by external fields,
\textcolor{blue} { Carbon {\bf143}, 14 (2019).}
\bibitem{K. Nakada} K. Nakada, M. Fujita, G. Dresselhaus, M. S. Dresselhaus, Edge state in graphene ribbons: Nanometer size effect and edge shape dependence,
\textcolor{blue} {Phys. Rev. B {\bf54}, 17954 (1996).}
\bibitem{Y.-W. Son} Y.-W. Son, M. L. Cohen and S. G. Louie, Energy Gaps in Graphene Nanoribbons,
\textcolor{blue} {Phys. Rev. Lett. {\bf97}, 216803 (2006).}
\bibitem{V. Barone} V. Barone, Oded Hod, and Gustavo E. Scuseria, Electronic Structure and Stability of Semiconducting Graphene Nanoribbons,
\textcolor{blue} { Phys. Rev. Lett. {\bf100}, 206802 (2008).}
\bibitem{Zhou} Benhu Zhou, Xiongwen Chen, Benliang Zhou, Kai-He Ding and Guanghui Zhou, Spin-dependent transport for armchair-edge graphene nanoribbons between ferromagnetic leads,
\textcolor{blue} { J. Phys.: Condens. Matter {\bf23}, 135304 (2011).}
\bibitem{Vy Tran} Vy Tran and Li Yang, Scaling laws for the band gap and optical response of phosphorene nanoribbons,
\textcolor{blue} { Phys. Rev. B {\bf89}, 245407 (2014).}
\bibitem{A. Carvalho} A. Carvalho, A. S. Rodin, and A. H. Castro Neto, Phosphorene nanoribbons,
\textcolor{blue} { Europhys. Lett. {\bf108}, 47005 (2014).}
\bibitem{W. F. Li} W. F. Li, G. Zhang, and Y. W. Zhang, Electronic Properties of Edge-Hydrogenated Phosphorene Nanoribbons: A First-Principles Study,
\textcolor{blue} { J. Phys. Chem. C {\bf118}, 22368 (2014).}
\bibitem{Y. F. Li} Y. F. Li, Z. Zhou, S. B. Zhang, and Z. F. Chen, MoS$_2$ Nanoribbons: High Stability and Unusual Electronic and Magnetic Properties,
\textcolor{blue} { J. Am. Chem. Soc. {\bf130}, 16739 (2008).}
\bibitem{Q. Y. Wu} Q. Y. Wu, L. Shen, M. Yang, Y. Q. Cai, Z. G. Huang, and Y. P. Feng, Electronic and transport properties of phosphorene nanoribbons,
\textcolor{blue} { Phys. Rev. B {\bf92}, 035436 (2015).}
\bibitem{S. Fathipour} S. Fathipour, M. Remskar, A. Varlec, A. Ajoy, R. Yan, S. Vishwanath, S. Rouvimov, W. S. Hwang, H. G. Xing, D. Jena, and A. Seabaugh, Synthesized multiwall MoS$_2$ nanotube and nanoribbon field-effect transistors,
\textcolor{blue} { Appl. Phys. Lett.  {\bf106}, 022114  (2015).}
\bibitem{V. Passi} V. Passi, A. Gahoi, B. V. Senkovskiy, D. Haberer, F. R. Fischer, A. Gruneis, and Max C. Lemme, Field-Effect Transistors Based on Networks of Highly Aligned, Chemically Synthesized N = 7 Armchair Graphene Nanoribbons,
\textcolor{blue} { ACS Appl. Mater. Inter. {\bf10}, 9900 (2018).}
\bibitem{Congxin Xia} C. X. Xia, L. Z. Fang, W. Q. Xiong, T. X. Wang, S. Y. Wei, Y. Jia, Rectification effects of C$_3$N nanoribbons-based Schottky junctions,
\textcolor{blue} { Carbon. {\bf141}, 363 (2019).}
\bibitem{M. Brandbyge} M. Brandbyge, J.L. Mozos, P. Ordejon, J. Taylor, K. Stokbro, Density-functional method for nonequilibrium electron transport,
\textcolor{blue} { Phys. Rev. B {\bf65}, 165401 (2002).}
\bibitem{J. Taylor} J. Taylor, H. Guo, J. Wang, Ab initio modeling of quantum transport properties of molecular electronic devices,
\textcolor{blue} { Phys. Rev. B {\bf63}, 245407 (2001).}
\bibitem{G. Kresse} G. Kresse, J. Furthmuller, Efficient iterative schemes for ab initio total-energy calculations using a plane-wave basis set,
\textcolor{blue} { Phys. Rev. B {\bf54}, 11169 (1996).}
\bibitem{Soler J. M} J. M. Soler, E. Artacho, J. D. Gale, A. Garcia, J. Junquera, P. Ordejon, and D. Sanchez-Portal, The SIESTA method for ab initio order-N materials simulation,
\textcolor{blue} {J. Phys. Condens. Matter {\bf14}, 2745 (2002).}
\bibitem{Monkhorst H. J} H. J. Monkhorst and J. D. Pack, Special points for Brillonin-zone integrations,
\textcolor{blue} {Phys. Rev. B {\bf13}, 5188 (1976).}
\bibitem{H. M. Pastawski} H. M. Pastawski, Classical anti quantum transport from generalized Landauer-Buttiker equations,
\textcolor{blue} { Phys. Rev. B {\bf44}, 6329 (1991).}
\bibitem{S. Dutta} S. Dutta, A. K. Manna and S. K. Pati, Intrinsic Half-Metallicity in Modified Graphene Nanoribbons,
\textcolor{blue} { Phys. Rev. Lett. {\bf102}, 096601 (2009).}
\bibitem{Fan} Z. Q. Fan, W. Y. Sun, X. W. Jiang, J. W. Luo, S. S. Li, Two dimensional Schottky contact structure based on in-plane zigzag phosphorene nanoribbon,
\textcolor{blue} {Org. Electron. {\bf44}, 20 (2017).}
\bibitem{W. X. Zhang} W. X. Zhang, C. Basaran and T. Ragab, Impact of geometry on transport properties of armchair graphene nanoribbon heterojunction,
\textcolor{blue} {Carbon {\bf124}, 422 (2017).}

\end{references}
\end{document}